\documentstyle{article}
\title{\LARGE The wave function of the modified space time manifold}
\author{A.~N.~Leznov\thanks{ Universidad Autonoma del Estado de Morelos, CCICAp,Cuernavaca, Mexico}} \date{}

\newcommand{\rig}[2]{\stackrel{#2\rightarrow}{#1}}

\begin{document}
\maketitle

\maketitle

\begin{abstract}
In the present paper it is considered space time manifold with four commutative coordinates with Poinc\'are group of motion but different from Poinc\
'are Minkowski space time of the modern physic.
\end{abstract}

\section{Introduction to introduction}

 The background of description of all events in our world is four dimensional space-time manifold and classical and quantum physical laws. The boundary is Plank constant $h$. Technically. on quantum  level we have  commutation relation between corresponding operators $[A,B]=ihC$. On classical level we deal with the  Poison brackets $\{A,B\}=C, \{A,B\}=\lim {[A,B]\over h},h\to 0$, however in both cases space-time manifold considered only on classical level. It is not dynamical system but only map  with coordinate system on which all events occurs.
In the present paper it will be shown how Poinc\'are-Minkowski space-time manifold may be generalized  to quantum-mechanical system. Here squared modulus  of  wave function $|f(X,P)|^2$ give possibility existent in in four-dimensional point  $X$ the objects with dynamical value of energy  momentum $P$. Possible this is some connection with description what is called by term the black matter. Under some conditions between $X,P$ quantum-mechanical description may descent to the classical level. This is exactly the sky of stars and galaxies  in the moment of $X_4=ct$. It is important to notice that the condition to observe a system as a classical one  is limited in the time. Condition of classical motion  may to be satisfied at some moment and cease this in the some other one. The stars are born the stars are died (black holes).

\section{Introduction}

Let us imagine for a moment that theory of relativity was discovered in the following logical pure theoretical logic manner. The physic of Newton is invariant (except of translations) to rotations $\rig{l}{}_i=\epsilon_{ijk} x_jp_k$ and transformation of Galileo $\rig{g}{}_i=tp_i$. These generators satisfy the following commutation relations
$$
[l_i,l_j]=\epsilon_{ijk}l_k, \quad [l_i,g_j]=\epsilon_{ijk}g_k, \quad [g_i,g_j]=0
$$
If one would try to modify this relation the obvious and simplest way is to add to the left hand side of the last equation a term proportional to ${1\over c^2}\epsilon_{ijk}l_k$. With dimension of parameter $c$ as velocity. Indeed $\rig{l}{}$ is dimensionless and $\rig{g}{}$ have inverse of velocity dimension. After this it would be possible to construct a theory invariant to new group of motion which will be different from mechanics of Newton only in the domain of velocity near to $c$. This would be the special relativity theory.

The same logic was used in attempts to construct quantum spaces.  However, the preliminary the goal here was to change the properties of Minkowski space only on microscopic space-time distances. But from dimension analysis it follows that all additional parameters arises in denominators exactly as $c$ in example above and to have correct results at least at distances of the solar system it is necessary to assume their values on the cosmic scale.

The most general form of the commutation relations between the elements of the quantum  four-dimensional space-time and its group of motion ($x$- coordinates, $p$- impulses, $F$- generators of Lorenz transformation, $I$ -"unity"" element) are the following ones \cite{LezKhr73}
$$
[p_i,x_j]=ih(g_{ij}I+{F_{ij}\over H}),\quad  [p_i,p_j]={ih\over L^2}F_{ij},
\quad [x_i,x_j]={ih\over M^2}F_{ij},
$$
\begin{equation}
[I,p_i]=ih(-{p_i\over H}+{x_i\over L^2}), \quad [I,x_i]=ih(-{p_i\over M^2}+
{x_i\over H}),\quad [I,F_{ij}]=0 \label{2}
\end{equation}
$$
[F_{ij},x_s]=ih(g_{is}x_j-g_{js}x_i),\quad [F_{ij},p_s]=ih(g_{is}p_j-g_{js}p_i)
$$
$$
[F_{ij},F_{sk}]=ih(g_{js}F_{ik}-g_{is}F_{jk}-g_{jk}F_{is}+g_{ik}F_{js})
$$
These relations are written under assuming that number of space-time elements is conserved and that Lorenz group is included into the main laws of the nature.
The second Kazimir operator of (\ref{2}) looks as
$$
K_2=-I^2+{(p)^2\over M^2}+{(x)^2\over L^2}-{(x p)+(p x)\over H}+
\sum_{i\leq j} (g_{i i}F_{i,j} g_{j,j}F_{i,j})({1\over H^2}-{1\over L^2M^2})\equiv
$$
\begin{equation}
-I^2+{(x-{L^2\over H}p)^2\over L^2}+({h^2\over H^2}-{h^2\over L^2M^2})
(-{L^2(p)^2\over h^2}+{\sum_{i\leq j} (g_{i i}F_{i,j} g_{j,j}F_{i,j})\over h^2})
\label{KAZ}
\end{equation}
And if we want in the limit have Poinc\'areMinkowski space-time with $I=1$  $K_2=-1$
commutation relations (\ref{2}) must be supplemented by some additional conditions which responsible for correct limit to usual Poinc\'areMinkowski space-time in the infinite limit of dimensional parameters. Such conditions looks as
$$
IF_{i,j}={x_j p_i-x_i p_j+p_i x_j-p_j x_i \over 2}
$$
 that in Poinc\'areMinkowski limit ($I\to 1$) represent relation between generators of Lorenz transformations and operators coordinates and momenta (more in detail see in next section).
The additional conditions above allow to fix definite representation of algebra
(\ref{2}) in the unique way \cite{LEZ04a},\cite{LEZ04b}.

To treat the problem on the classical level (\ref{2}) it is necessary rewrite with exchanging
$$
\lim {[A,B]\over ih}\to \{A,B\}
$$
(change all commutator of quantum theory on corresponding Poisson brackets) and consider
(\ref{2}) on the level of functional algebra \cite{Eis33}).

The commutation relations of the quantum space contain 3 dimensional parameters
of the dimension length $L$, the momenta$Mc\to M$,\footnote {We use the system units where the speed of light is equal to unity} and the action $H$. The
equalities of Jacoby  are satisfied for (\ref{2}). It should be stressed that the  signs
of $L^2,M^2$ are not required to be positive.
The limiting procedure $ M^2,H\to \infty$ one comes to the space of constant curvature, considered in connection with Column problem by
E.Schredinger \cite{Sch40}, $L^2,H\to \infty$ leads to quantum space of Snyder \cite{Sny47}, $H\to \infty$ leads to Yang quantum space \cite{Yan47}. Except of $L^2,M^2$ parameter dimension of action $H$ was introduced in  \cite{LezKhr73}.

The term quantum  space is not very adequate to this problem, because the modified classical dynamics may be considered here (also as electrodynamics, gravity theory and so on). More correct but  not so short this problem may be called as possible generalized manifold of the world and its group of motion.

In general (\ref{2}) is commutation relations one of real forms of six-dimensional rotation group. In the case $L^2\to \infty$ as it follows directly from (\ref{2}) commutation relations between $p_i,F_{i,j}$ exactly coincide with commutation relation of Poincare algebra and thus the last is the group of motion of the space-time manifold under consideration in this case.
Such version of quantum space was considered in \cite{LEZ08a}.

\section{Realization of quantum space on the base the usual Minkovski one}

Realization algebra (\ref{2}) in general case of arbitrary $(L^2,M^2,H)$ on the base Poinc\'are algebra was presented in \cite{Lez08b}.

As was mentioned above in the present paper will be considered  the most simple case of quantum space-time manifold with only one dimensional parameter of action $H$. In this case it is possible all calculations represent in a closed form. Algebra of space-time manifold looks in this case as
$$
[p_i,x_j]=ih(g_{ij}I+{F_{ij}\over H}),\quad  [p_i,p_j]=0,\quad [x_i,x_j]=0,
$$
\begin{equation}
[I,p_i]=-ih{p_i\over H}, \quad [I,x_i]=ih{x_i\over H}),\quad [I,F_{ij}]=0 \label{A3}
\end{equation}
$$
[F_{ij},x_s]=ih(g_{is}x_j-g_{js}x_i),\quad [F_{ij},p_s]=ih(g_{is}p_j-g_{js}p_i)
$$
$$
[F_{ij},F_{sk}]=ih(g_{js}F_{ik}-g_{is}F_{jk}-g_{jk}F_{is}+g_{ik}F_{js})
$$
This is space-time manifold with four commutative coordinates and Poinc\'are group of motion. The Lie algebra defined by (\ref{A3}) is Lie algebra of $O(2.4)$ noncompact orthogonal  group (equivalent to conformal one $SU(2.2)$)
The representation of conformal algebra it is necessary choose keeping in mind that in limiting case $H\to \infty$ generators of Poinc\'are algebra satisfy 15 additional relations
$$
F_{i,j}=x_i p_j-x_j p_i,(6),\quad \epsilon_{i,j,k,l}F_{i,j}x_k=0, (4),\quad \epsilon_{i,j,k,l}F_{i,j}p_k=0, (4),
$$
$$
\epsilon_{i,j,k,l}F_{i,j}\epsilon_{i,j,k,l}F_{k,l}=0, (1).
$$
Thus representation of modified space-time algebra must be chosen in such way that its generators in the limit $H\to \infty$ satisfy the above conditions. From generators of $O(6)$  algebra $O_{i,j}$ $1\leq i,j\leq 6$ it is possible to construct 15 quadratical ones operators  $\epsilon_{i,j,k,l,m,n}O_{k,l}O_{m ,n}=0$ $0_{5,i}=x_i,0_{6,i}=p_i,0_{5,6}=I$ which limit exactly coincide     with written above relations. The main among such constructed quadratical operators is is the following one
$$
IF_{i,j}={1\over 2}[(p_ix_j-p_jx_i)+(x_j p_i-x_i p_j)]=(p_ix_j-p_jx_i)+ih{F_{ji}\over H}
$$
$K_2$ in the case under consideration $\to \infty$ takes the form
$$
K_2=-I^2+{(x p)+(p x)\over H}+
\sum_{i\leq j} (g_{i i}F_{i,j} g_{j,j}F_{i,j}){1\over H^2}
$$
The form of realization quantum space is demand for its understanding some knowledge  from representation theory of semi-simple algebras \cite{LEZ04a},\cite{LEZ04b}. Here we present another form of realization (of course equivalent to previous one).  For checking of its validity it is sufficient to be acquainted with the first chapters of usual curse of quantum mechanic.

Let us look for operators of four dimensional coordinates and momenta in a form \cite{LezFed74} (really this paper was done only by I.A.Fedoseev and author was known that he is coauthor only after its publication).
\begin{equation}
\bar p_i=a p_i+b A_i,\quad \bar x_i=c p_i+d A_i,\quad A_i=\rho x_i+[{x^2+1\over 2}p_i-x_i(x p)]
\label{DEF}
\end{equation}
where $\bar x,\bar p$ coordinates and momenta of real space-time and $p,x$ non
physical the same of Minkowski space with commutation relations $[p_i,x_j]=g_{i,j},g_{44}=1,g_{\alpha,\alpha}=-1$ and
$x^2=x_4^2-(\rig{x}{})^2,(x p)=x_4 p_4-(\rig{x}{}\rig{p}{})$, $a,b,c,d,\rho$
arbitrary parameters which will be defined below.

In what follows all operators above will be considered dividing on $ih$ and thus the obtained finally values will be necessary multiply on this value.

From definitions above the following commutation relations take place
$$
[A_i,A_j]=x_i p_j-x_j p_i\equiv F_{i,j},\quad [p_i,A_j]=g_{i,j}(\rho-(x p))+F_{i,j},\quad [(x p),x_i]=x_i,
$$
$$
[(x p),p_i]=-p_i,\quad [(x p),A_i]=A_i-p_i
$$
Commutation relations $[\bar p_i,\bar p_j]=0, [\bar x_i,\bar x_j]=0$ equivalent correspondingly to $2ab+b^2=0,2cd+d^2=0$  with nontrivial pair solutions $d=0, b+2a=0$ or $ b=0, d+2c=0$. In what follows it will be used the first one. Thus $\bar x_i=cp_i,
\bar p_i=b[{x^2\over 2}p_i+x_i(\rho-(x p))]$

Now
$$
[\bar p_i,\bar x_i]=-bcg_{i,j}(\rho-(x p))+bc(x_ip_j-x_jp_i)=g_{i,j}\bar I+{1\over H}\bar F_{i,j}
$$
From which we obtain
$$
-bc={1\over H},\quad I={1\over H}(\rho-(x p)),\quad \bar F_{i,j}=F_{i,j}=x_jp_i-x_ip_j
$$
At last
$$
[\bar I,\bar p_i]={b\over H}[(\rho-(x p)),({x^2\over 2}p_i+x_i(\rho-(x p)))]=-{1\over H}b({x^2\over 2}p_i+x_i(\rho-(x p))=-{1\over H}\bar p_i
$$
$$
[\bar I,\bar x_i]={c\over H}[(\rho-(x p)),p_i]={c\over H}p_i= {1\over H}\bar x_i
$$

Now it is necessary to check additional condition. We use transform form of it
$$
(\bar p_i\bar x_j-\bar p_j\bar x_i)+ih{F_{ji}\over H}={F_{ji}\over H}+bc({x^2\over 2}p_i+x_i(\rho-(x p)) p_j-
{x^2\over 2}p_j+x_j(\rho-(x p)) p_i)=
$$
$$
{F_{ji}\over H}-bc(\rho+1-(x p))F_{ij}=\bar I \bar F_{ij},\quad bc=-{1\over H} (!)
$$

Thus algebra (\ref{A3}) is realized in a form
$$
\bar p_i=ihb[{x^2\over 2}p_i+x_i(\rho-(x p))],\quad \bar x_i=ihcp_i,\quad \bar I={ih\over H}(\rho-(x p)),\quad \bar F_{ij}=ih(x_jp_i-x_ip_j)
$$

Operator Kazimira calculated by the same manner lead to result $K_2={h^2\over H^2}\rho(\rho+4)$. Condition that all operators of the physical values under consideration are hermitian ones (that representation is unitary) we will consider on example of $\bar I$ operator. We have consequently
$$
(\bar I)^{Her}=-{ih\over H}(\rho*+(p x))={ih\over H}(-(\rho*+4)-(xp))=\bar I
$$
Lead to $ \rho*+4+\rho=0, \rho=-2+i\nu$. Equality of the second Kazimir operator to unity determine $\nu=\pm\sqrt{{H^2\over h^2}-4},\rho=-2\pm i\sqrt{{H^2\over h^2}-4}$.

\section{Free motion in classical domain}

In Poinc\'are-Minkovski space-time this is the motion with constant velocity by the linear lines. In considered modified space-time equation of the free motion are the following one (solution in the general case see in \cite{Lez07})
\begin{equation}
1=I^2-{2(px)\over H}+{f^2-l^2\over H^2},\quad I\vec f=x_4\vec p-p_4 \vec x,\quad I\vec l=[\vec x x \vec p] \label{4}
\end{equation}
From which it follows directly $ \vec l=-{1\over p_4}[\vec f x \vec p],\quad {f^2-l^2\over H^2}={f^2\over H^2}{m^2\over p_4^2}+{(pf)^2\over p_4^2H^2},\quad m^2 =p_4^2-p^2$ These results substituting in the first equation of (\ref{4}) lead to
$$
1=I^2-2{x_4p_4\over H}+2{\vec p (x_4\vec p-I\vec f)\over p_4 H}+{f^2\over H^2}{m^2\over p_4^2}+{(pf)^2\over p_4^2H^2}=
$$
$$
(I-{(pf)\over p_4H})^2+{m^2\over Hp_4}(-2x_4+{f^2\over H p_4})=1
$$
From the last result it follows that classical consideration is impossible only if the the time satisfying the
condition $1+{m^2\over Hp_4}(2x_4-{f^2\over H p_4})\leq 0$. Let us investigate from what space-time point is impossible classical consideration.
On the boundary we have
$$
{Hp_4\over m^2}+(2x_4-{f^2\over H p_4}=0,\quad I={(pf)\over p_4H},\quad {(pf)^2\over p_4H}=p_4(XP)-x_4P^2,\quad {(pf)^2\over (p_4H)^2}f^2=
$$
$$
-p_4^2 X^2+2x_4p_4 (XP)-p_4^2 X^2,\quad (x_4P^2++{Hp_4\over 2})^2=p_4^2(X^2P^2+{H^2\over 4}+H(XP))
$$
In the limit $H\to \infty$ the last limitation pass to equality.

Some conclusion to this section, From consideration of classical laws o motion in modified world it follows that classically observable
object arise in some moment in some space point $X$ with energy-momenta $U$. Question what was before? Only one possible answer is that it was on quantum (non classically observable) level and only in the moment of its born the wave function of it take quasi classical form.
In the next section it will be investigated the possible form of the wave function.

\section{Matrix elements}

We use symbols $X_i,P_i$ for coordinates and impulses of real physical world and  $x_i,p_i$ the same of Minkowski one.

Operators of coordinates are commutative and thus matrix element $<X|x>$ is solution of the equation $X_k <X|x>=\bar x_k <X|x>=ihcp_k <X|x>$.
Finally $<X|x>=e^{{(xX)\over (ihc}}$.
Matrix element $<P|x>\equiv r$ is solution of the system (operators $P_k=\bar p_k$ are commutative ones!)
$$
P_k r=\bar p_k=ihb[{x^2\over 2}p_k+x_k(\rho-(x p))] r,\quad {P_k\over ihb}-\rho x_k={x^2\over 2}p_k-x_k(x p)]\ln r
$$
After multiplication the last equation on $x_k$ on the left with consequent summarizing we obtain $ (x p)\ln r=2\rho-2{(xP)\over ihb x^2}$
Substituting this result to initial equation after trivial integration we obtain finally $ r=<P|x>=e^{{2(xP)\over ih b x^2}+\rho \ln x^2}$

At last for matrix element in real space we have
\begin{equation}
<P|X>=\int d^4 x <P|x><X|x>^*=\int d^4 x e^{-i({(2(xP)\over b x^2}0{(xX)\over c})+(-2+i\nu) \ln x^2} \label{4}
\end{equation}
$-2+i\nu=\rho,\nu=\sqrt{({H\over h})^2-4}.$ Parameters ${b\over 2},-c$  will be included into $P,X$ correspondingly.

\subsubsection{Some trivial but important properties of $<P|X>$}

$<P|X>$ is invariant with respect to exchange $X_i\to -X_i, P_i\to -P_i$ for arbitrary $i$.
$<P|X>(X,P)=(a^2)^{\nu}<P|X>(aX,{P\over a})$ for arbitrary real $a$
$<P|X>(X,P)=<P|X>^*(X,P)$

These properties very simple to proof exchanging variables in initial integral $ x_i\to -x_i,x_i\to ax_i, x_i\to {x_i\over x^2}$

\subsection{$0\leq X^2=X_4^2-(\vec X)^2$}

By corresponding Lorenz transformation four dimensional vector $ X_4,\vec X$ under condition $0\leq X^2$ may represented to form $(X_4,0,0,0)$ and after this other Vector $P$ only with the three dimensional rotation may be transform  to a form $P_4.P_3,0,0)$. Thus after over going to spherical coordinates in three dimensional space we obtain
$$
<P|X>=\int dx_4 r^2 d r\sin \theta d\theta d \phi e^{-i({((x_4P_4-r cos \theta P_3)\over  x^2}+(x_4X_4))+(-2+i\nu) \ln |x^2|}=
$$
$$
 {2\pi\over iP_3} \int dx_4 r x^2 d r [e^{-i({((x_4P_4-r P_3)\over  x^2}+(x_4X_4))+(-2+i\nu) \ln |x^2|}-(P_3\to -P_3)]
$$
In the two dimensional integral above let us exchange  variables $x_4=r {\cos \sigma\over \sin \sigma}$, $x^2=r^2{\cos 2\sigma\over\sin^2\sigma}$
 All common factors will be taken into account in the end of calculations.
$$
\int_0^{\pi} d \sigma \int_0^{\infty} d r{r^2\over \sin^2\sigma}{\sin^2(\sigma)\over r^2\cos 2 \sigma}(r^2|{\cos 2 \sigma\over\sin^2(\sigma)}|)^{i\nu}
$$
$$
[e^{-i{(\cos\sigma P_4+\sin \sigma P_3)\over  r{\cos 2 \sigma\over\sin(\sigma)}}+(r {\cos \sigma\over \sin \sigma}X_4}
-(P_3\to -P_3)]
$$

After exchanging ${r\over \sin \sigma}\to r,0\leq \sin \sigma$ The last integral looks as
$$
\int_0^{\pi}{\sin(\sigma)\over \cos 2 \sigma} d \sigma \int_0^{\infty} dr (r^2|\cos 2 \sigma|)^{i\nu}
$$
$$
[e^{-i{(\cos\sigma P_4-\sin \sigma P_3)\over b r\cos 2 \sigma}+c(r \cos \sigma X_4}-(P_3\to -P_3)]
$$
Integration over $\sigma$ will be divided on 4 domains $(0,{\pi\over 4}),({\pi\over 4},{\pi\over 2}),({\pi\over 2},{3\pi\over 4}).({3\pi\over 4},\pi)$
In the first one $0\leq cos 2\sigma$ and after exchanging $r \sqrt{cos 2\sigma}\to r,\quad {cos \sigma\over \sqrt{cos 2\sigma}}\to \cosh u,\quad
{\sin \sigma\over \sqrt{cos 2\sigma}}\to \sinh u,\quad d \cosh u={\sin(\sigma)\over \cos^{3\over 2} 2 \sigma} d\sigma,\quad 0\leq u \leq \infty$
and integral on the first domain looks as
$$
\int_0^{\infty} d\cosh u \int_0^{\infty}(r^2)^{i\nu} dr [e^{-i({P_4\cosh u-P_3\sinh u\over b r}+c\cosh u X_4)}-(P_3\to -P_3)]
$$
Integration on 4 domain the following exchanging of variables $r \sqrt{cos 2\sigma}\to r,\quad {cos \sigma\over \sqrt{cos 2\sigma}}\to
-\cosh u,\quad {\sin \sigma\over \sqrt{cos 2\sigma}}\to \sinh u,\quad d \cosh u=-{\sin(\sigma)\over \cos^{3\over 2} 2 \sigma} d\sigma,\quad 0\leq u \leq \infty$ ($\cos \sigma$ in this domain is negative and integral on $u$ is from $\infty$ up to zero) lead to result
$$
\int_0^{\infty} d\cosh u \int_0^{\infty}(r^2)^{i\nu} dr [e^{i({P_4\cosh u+P_3\sinh u\over b r}+c\cosh u X_4)}-(P_3\to -P_3)]
$$
In both cases we integrals of the form $ \int_0^{\infty} de^{u}W(e^{u},e^{-u})+\int_0^{\infty} de^{-u}W(e^{u},e^{-u})$. After exchanging of variables $e^{u}\to T$ in the first integral and $e^{-u}\to T$ in the second one we obtain $\int_1^{\infty} dT W(T,T^{-1})+\int_1^0 dT W(T^{-1},T)$. But
$W$ function is antisymmetric  with respect exchanging $T\to T^{-1}$ and thus integral on $1,4$ domains looks as
$$
\int_0^{\infty} dT \int_0^{\infty}(r^2)^{i\nu} dr [e^{-{i\over 2}{(P_4(T+T^{-1})-P_3(T-T^{-1})\over b r}+c(T+T^{-1}) X_4)}+
$$
$$
e^{{i\over 2}{(P_4(T+T^{-1}) +P_3(T-T^{-1})\over b r}+c(T+T^{-1}) X_4)}-(P_3\to -P_3)]=
$$
$$
2i\int_0^{\infty} dT \int_0^{\infty}(r^2)^{i\nu} dr[\sin({1\over 2}({P_-T+P_+T^{-1}\over b r}+c(T+T^{-1}) X_4)-
$$
$$
\sin{1\over 2} ({P_+T+P_-T^{-1}\over b r}+c(T+T^{-1}) X_4)].\quad P_{\pm}\equiv P_4\pm P_3
$$

In $(2,3)$ domains $cos 2\sigma\leq 0$ and by corresponding exchange of variables $r \sqrt{-cos 2\sigma}\to r,\quad {cos \sigma\over \sqrt{cos -2\sigma}}\to\pm\sinh u,\quad {\sin \sigma\over \sqrt{-cos 2\sigma}}\to \cosh u,\quad d \sinh u=\mp{\sin(\sigma)\over \cos^{3\over 2} -2 \sigma} d\sigma,\quad 0\leq \pm u \leq \infty$ ($\cos \sigma$ positive in second domain and negative in the third) lead to the result
$$
-\int_0^{\infty} d\sinh u \int_0^{\infty}(r^2)^{i\nu} dr [e^{{i\over 2}({P_4\sinh u-P_3\cosh\over b r}- cr\sinh u X_4)}+
$$
$$
e^{-({P_4\sinh u+P_3\cosh u\over b r}-cr\sinh u X_4)}-(P_3\to -P_3)]
$$
As in the previous case this integral rewritten in variable $T,r$ takes the form
$$
-\int_0^{\infty} d T \int_0^{\infty}(r^2)^{i\nu} dr[e^{{i\over 2}({P_4(T-T^{-1}) u-P_3(T+T^{-1})\over b r}- c(T-T^{-1}) rX_4)}+
$$
$$
e^{-{i\over 2}({P_4(T-T^{-1})+P_3(T+T^{-1})\over b r}-c(T-T^{-1})r X_4)}-(P_3\to -P_3)]=
$$
$$
-\int_0^{\infty} d T \int_0^{\infty}(r^2)^{i\nu} dr[\sin {1\over 2}({P_-T-P_+T^{-1}\over b r}- c(T-T^{-1}) r X_4)-
$$
$$
\sin {1\over 2}({P_+T-P_-T^{-1}\over b r}- c(T-T^{-1}) r X_4)]
$$
And thus integral on all four domains is equal to
$$
\int_0^{\infty} d T \int_0^{\infty}(r^2)^{i\nu} dr [\sin {1\over 2}({P_+T\over b r}+c{r\over T})X_4)\cos{1\over 2}({P_-\over b Tr}+ cTr X_4)-
$$
$$
\sin {1\over 2}({P_-T\over b r}+c{r\over T})X_4)\cos{1\over 2}({P_+\over b Tr}+ cTr X_4)]
$$
At last from variables $T,r$ pass to variables $x={T\over r},y=Tr)$ $r^2-{y\over x},T^2=xy$ ,$J{\pmatrix{x & y \                                                                                                       T & r \cr}}=2{T\over r}=2x$ with the finally result
$$
{1\over P_3}\int_0^{\infty} d y \int_0^{\infty}({y\over x})^{i\nu} {dx\over x} [\cos {1\over 2}(P_-x+{X_4\over x})\sin{1\over 2}({P_+\over b y}+ cy X_4)-
$$
$$
\cos {1\over 2}({P_+x\over b}+c{X_4\over x})\sin{1\over 2}({P_-\over b y}+ cy X_4)]
$$
By direct calculations it possible to check that this expression is invariant with respect to transformation
$X_3\to -X_3,P_3\to -P_3,P_+\to P_-,P_-\to P_+$,$X_4\to -X_4,P_4\to -P_4,P_+\to -P_-,P_-\to -P_+$.

\begin{equation}
\int \int dx dy (xy)^{i\nu} x^{-1}[ e^{-i{1\over 2}(X_4 x+{P_+\over x})}e^{i{1\over 2}(X_4 y+{P_-\over y})}-e^{-i{1\over 2}(X_4 x+{P_-\over x})}e^{i{1\over 2}(X_4 y+{P_+\over y})}
$$
$$
-e^{i{1\over 2}(X_4 x+{P_-\over x})}e^{-i{1\over 2}(X_4 y+{P_+\over y})}+e^{i{1\over 2}(X_4 x+{P_+\over x})}e^{-i{1\over 2}(X_4 y+{P_-\over y})}
$$
$$
+e^{-i{1\over 2}(X_4 x+{P_{-}\over x})}e^{-i{1\over 2}(X_4 y+{P_{+}\over y})}-e^{-i{1\over 2}(X_4 x+{P_{+}\over x})}e^{-i{1\over 2}(X_4 y+{P_-\over y})}
$$
\label{6}
$$
 - e^{i{1\over 2}(X_4 x+{P_+\over x})}e^{i{1\over 2}(X_4 y+{P_-\over y})}+e^{i{1\over 2}(X_4 x+{P_-\over x})}e^{i{1\over 2}(X_4 y+{P_+\over y})}]=
 \end{equation}
$$
 {1\over P_3}\int_0^{\infty}d x \int_0^{\infty} dy (xy)^{i\nu} x^{-1}
 $$
 $$
 [\sin {1\over 2} (X_4 x+{P_+\over x})\cos {1\over 2}(X_4 y+{P_-\over y})-\sin {1\over 2}(X_4 x+{P_- \over x})\cos {1\over 2}(X_4 y+{P_+\over y})]
 $$

\subsection{$ X^2=X_4^2-(\vec X)^2 \leq 0$}

 In this case two four dimensional vectors may be by Lorenz transformation represent in a form $X=(0,X_3,0,0), P=(P_4,P_3.0.0)$. In three dimensional spherical system coordinates after integration On $\theta$ matrix element $<P|X>$ looks as
 $$
 <P|X>i\int_{-\infty}^{\infty} dx_4\int_0^{\infty} {r d r (|x^2|)^{i\nu}\over x^2(P_3+X_3 x^2)}
 [ e^{-i({((x_4P_4-r P_3)\over b x^2}-c(rX_3))}-(X_3\to -X_3, P_3\to -P_3)]
$$
 After introduction the the same new variables as in the case $0\leq X^2$ and fullfil all the same transformations with corresponding exchanging of variables we obtain finally variable $x_4=r {\cos \sigma\over \sin \sigma}$, $x^2=r^2{\cos 2 \sigma\over\sin^2(\sigma)}$ integral above takes the4 form (up to numerical factors)
 $$
 \int_0^{\pi} d\sigma {1\over \cos 2 \sigma} \int_0^{\infty} {d r (|r^2{\cos 2 \sigma\over\sin^2(\sigma)}|)^{i\nu}\over 2(P_3+X_3 r^2{\cos 2 \sigma\over\sin^2(\sigma)})}
 [ e^{-i({((x_4P_4+r P_3)\over b x^2}+c(rX_3))}-(X_3\to -X_3, P_3\to -P_3)]=
 $$
 \begin{equation}
 \int_0^{\infty} d y \int_0^{\infty}({y\over x})^{i\nu} {dx\over x}\label{RAL}
 \end{equation}
$$
{1\over (P_3+X_3 {y\over x})}[\sin (-{X_3\over x}+P_+x+X_3 y+{P_-\over y})-\sin ({X_3\over x}+P_-x-X_3 y+{P_+\over y})]-
$$
$$
{1\over (P_3-X_3 {y\over x})}[\sin ({X_3\over x}+P_-x+X_3 y-{P_+\over y})+\sin ({X_3\over x}-P_+x+X_3 y+{P_-\over y})]\}
$$
To transform this integral let us cambia $(x,y)\to (x,z={y\over x})$. The goal of this transformation no do concrete calculations but represent the result above in equivalent form with later integration. The following functions will be used by the way.
$\theta (x)=1  (0\leq x), =0 (x\leq 0). \theta_x (x)=\delta (x)$ -delta function of Dirac
 The first  part integral on $x$ looks as
$$
{1\over (P_3+X_3z)}\int_0^{\infty} {d x\over x}[{\frac{\partial}{\partial P_+}}\cos ({{P_-\over z}-X_3\over x}+x(P_++X_3z)) -{\frac{\partial}{\partial P_-}}\cos ({{P_+\over z}+X_3\over x}+x(P_--X_3 z))]
$$
In connection of mathematics results below  integration on $x$ lead to
$$
{1\over (P_3+X_3 z)}[({\frac{\partial}{\partial P_+}}-{\frac{\partial}{\partial P_-}})(\theta (P_-zX_3)\theta (P_++zX_3)+
(\theta (-(P_-zX_3))\theta -(P_++zX_3)+........)
$$
$$
Y_0(\sqrt{{(P_-zX_3)(P_++zX_3)\over z}}
$$
Dots in above line means all other possibilities of common signs factors  $(P_-zX_3)(P_++zX_3)$. Consider at first result of differentiation Bessel $Y_0$
$$
{1\over (P_3+X_3 z)}{1\over 2}[\sqrt{{(P_--zX_3)\over(P_++zX_3)z}}-\sqrt{{(P_++zX_3)\over (P_--zX_3)z}}]\dot Y_0=-\sqrt{{1\over (P_++zX_3)(P_--zX_3)z}}Y_0=
$$
$$
-{1\over 2P_4}[\sqrt{{(P_-zX_3)\over(P_++zX_3)z}}+\sqrt{{(P_++zX_3)\over (P_-zX_3)z}}]\dot Y_0=
$$
$$
-{1\over P_4}({\frac{\partial}{\partial P_+}}+{\frac{\partial}{\partial P_-}})Y_0
$$
Now let us consider differentiation factor with $\theta$ functions
$$
{1\over (P_3+X_3 z)}[\theta (P_-zX_3)\delta (P_++zX_3)+
-(\theta -(P_-zX_3)\delta (P_++zX_3)-
$$
$$
\delta (P_-zX_3)\theta (P_++zX_3)+(\delta (P_-zX_3)\theta -(P_++zX_3))Y_0(0)=
$$
$$
-{1\over P_4}\delta (P_++z X_3)(\theta P_4-\theta -P_4)--{1\over P_4}\delta (P_+-zX_3)(\theta P_4-\theta -P_4)=
$$
$$
-{1\over P_4}-{1\over P_4}({\frac{\partial}{\partial P_+}}+{\frac{\partial}{\partial P_-}})(\theta (P_-zX_3)\theta (P_++zX_3)+(\theta (-(P_-zX_3))\theta -(P_++zX_3)
$$
Coming back step by step to (\ref{RAL}) we obtain instead of it
\begin{equation}
 -{1\over P_4}\int_0^{\infty} d y \int_0^{\infty}({y\over x})^{i\nu} {dx\over x}\label{RAL1}
 \end{equation}
$$
[\sin (-{X_3\over x}+P_+x+X_3 y+{P_-\over y})+\sin ({X_3\over x}+P_-x-X_3 y+{P_+\over y})-
$$
$$
\sin ({X_3\over x}+P_-x+X_3 y-{P_+\over y})+\sin ({X_3\over x}-P_+x+X_3 y+{P_-\over y})]=
$$
\begin{equation}
 -{1\over P_4}\int_0^{\infty} d y \int_0^{\infty}({y\over x})^{i\nu} {dx\over x}\label{RAL1}
 \end{equation}
 $$
 \sin {1\over 2}(-X_3 y+{P_+\over y})\cos {1\over 2}({X_3\over x}+P_-x)+\sin {1\over 2}(X_3 y+{P_-\over y})\cos {1\over 2}({X_3\over x}-P_+x)
 $$

\subsection{Necessary Mathematics}

All results it is possible well known text book Yu 
For  Bessel functions $I,K$ and  $ J,Y$ correspondingly take place recurrent relations
$$
I_{\mu-1}(x)-I_{\mu+1}(x)={2\mu\over x}I_{\mu}(x),\quad Y_{\mu+1}(x)+Y_{\mu-1}(x)={2\mu\over x}Y_{\mu}(x)
$$

With the help of definition $K_{\mu},J_{\mu}$ Bessel functions via $I_{\mu},Y_{\mu}$ ones
$$
K_{\mu}={\pi\over 2}{I_{-\mu}-I_{\mu}\over \sin \pi\mu}. \quad J_{\mu}={\cos(\mu \pi)Y_{-\mu}-Y_{\mu}\over \sin \pi\mu}
$$
below present necessary for further calculations integrals
$$
\int_0^{\infty} dx x^{\mu-1} \cos(a(x-{b^2\over x}))=\pi b^{\mu}{I_{-\mu}(2ab)-I_{\mu}(2ab)\over 2\sin{\mu \pi\over 2}},
$$
$$
\int_0^{\infty} dx x^{\mu-1} \sin(a(x-{b^2\over x}))=\pi b^{\mu} {\mu}{I_{-\mu}(2ab)-I_{\mu}(2ab)\over 2\cos{\mu \pi\over 2}},
$$
$$
\int_0^{\infty} dx x^{\mu-1} \sin(a(x+{b^2\over x}))=\pi b^{\mu}{Y_{-\mu}-Y_{\mu}\over 2\sin{\mu \pi\over 2}}
$$
$$
\int_0^{\infty} dx x^{\mu-1} \cos(a(x+{b^2\over x}))=-\pi b^{\mu}{Y_{-\mu}+Y_{\mu}\over 2\cos{\mu \pi\over 2}}
$$

\subsection{Direct calculations}

In all calculations under integration on $x$ $\mu=i \nu$ in the same on $y$ $\mu=i \nu+1$

\subsubsection{$0\leq X^2$}

In this case as it was obtained above (\ref{6})
$$
<X|P>=\int \int dx dy (xy)^{i\nu} x^{-1}=
$$
$$
 {1\over P_+-P_-}\int_0^{\infty}d x \int_0^{\infty} dy (xy)^{i\nu} x^{-1}[\sin {1\over 2} (X_4 x+{P_+\over x})\cos {1\over 2}(X_4 y+{P_-\over y})-
 $$
 $$
 \sin {1\over 2}(X_4 x+{P_- \over x})\cos {1\over 2}(X_4 y+{P_+\over y})]
 $$
In all calculations under integration on $x$ $\mu=i \nu$ in the same on $y$ $\mu=i \nu+1$. $\mu$ parameters on integrals of $\cos,\sin$ in mathematical subsection.
The function is invariant with transformation  and $ P_3\to-P_3,P_-\to P_+,P_-\to P_+$. This fact allow represent integral in 3 parts
$$
[\theta(X_4)\theta(P_+)\theta(P_-)+\theta(-X_4)\theta(-P_+)\theta(-P_-)]F_1=
$$
$$
\theta(P^2)[\theta(X_4)\theta(P_4)+\theta(-X_4)\theta(-P_4))]F_1=\theta(P^2)\theta(X_4P_4)F_1
$$
$$
[\theta(-X_4)\theta(P_+)\theta(P_+)+\theta(X_4)\theta(-P_+)\theta(-P_+)]F_2=
$$
$$
\theta(P^2)[\theta(-X_4)\theta(P_4)+\theta(X_4)\theta(-P_4))]F_2=\theta(P^2)\theta(-X_4P_4)F_2
$$
$$
[\theta(X_4)(\theta(P_+)\theta(-P_-)+\theta(P_-)\theta(-P_+)+\theta(-X_4)(\theta(P_+)\theta(-P_-)+\theta(-P_+)\theta(P_-))]F_3=
\theta(-P_+P_-)F_3
$$
We clarify all calculations on example of $F_3$ function. In the case $0\leq P_+,P_-\leq 0,0\leq X_4$ the function under integral looks as
$$
{1\over |P_+|+|P_-|}[\sin {1\over 2}|X_4|(x+{|P_+|\over x|X_4|})\cos {1\over 2}|X_4|(x-{|P_-|\over x|X_4|})-\sin {1\over 2}|X_4|(x-{|P_-|\over x|X_4|})\cos {1\over 2}|X_4|(x+{|P_+|\over x|X_4|})
$$

Thus in all formulae above $a={1\over 2}|X_4|,b=\sqrt {{|P_{\pm}|\over |X_4|}}, 2ab=\sqrt {|P_{\pm}||X_4|}\equiv z_{\pm}$.
Using formulae mathematical subsection we obtain for $F_2$
$$
\pi ({|P_{+}|\over |X_4|})^{\mu\over 2}{Y_{-\mu}(z_+)-Y_{\mu}(z_+)\over 2\sin {\mu\pi\over 2}}\pi ({|P_{-}|\over |X_4|})^{\mu+1\over 2}{I_{-(\mu+1)}(z_-)-I_{(\mu+1)}(z_-)\over 2\sin {(\mu+1)\pi\over 2}}[
$$
$$
\pi ({|P_{-}|\over |X_4|})^{\mu\over 2}{Y_{-\mu}(z_-)-Y_{\mu}(z_-)\over 2\cos {\mu\pi\over 2}}\pi ({|P_{+}|\over |X_4|})^{\mu+1\over 2}[-{I_{-(\mu+1)}(z_+)+Y_{(\mu+1)}(z_+)\over 2\cos {(\mu+1)\pi\over 2}}]=
$$
$$
{\pi^2\over 4\sin {\mu\pi\over 2}\cos {\mu\pi\over 2}}({|P_{+}||P_{-}|\over |X_4|^2})^{\mu+1\over 2}|X_4|^2[{Y_{-\mu}(z_+)-Y_{\mu}(z_+)\over z_+}(I_{-(\mu+1)}(z_-)-I_{(\mu+1})-
$$
$$
({I_{-\mu}(z_-)-I_{\mu}(z_-)\over z_-})(Y_{-(\mu+1)}(z_+)+Y_{(\mu+1)}(z_+)]
$$
With the help of recurrent relations  for $Y,I$ we have
$$
{Y_{-\mu}(z_+)-Y_{\mu}(z_+)\over z_+}=-{1\over 2\mu}[Y_{-\mu+1}(z_+)+Y_{-\mu-1}(z_+)+Y_{\mu+1}(z_+)+Y_{\mu-1}(z_+)]
$$
$$
{I_{-\mu}(z_-)-I_{\mu}(z_+)\over z_-}=-{1\over 2\mu}[I_{-\mu-1}(z_-)-I_{-\mu+1}(z_-)+I_{\mu-1}(z_-)-I_{\mu+1}(z_-)]
$$
Finally we obtain
$$
F_2=-{\pi^2\over 4\mu\sin (\mu\pi)(|P_+|+|P_-|}[Y_{-\mu+1}(z_+)+Y_{\mu-1}(z_+))(I_{-(\mu+1)}(z_-)-I_{(\mu+1})(z_-))-
$$
$$
(Y_{(\mu+1)}(z_-)+Y_{(\mu+1)}(z_-)(I_{-(\mu+1)}(z_+)-I_{(\mu+1})(z_+))]({|P_{+}P_{-}|\over |X^2_4|})^{\mu+1\over 2}
$$
Absolutely by the same technique we obtain ( $0\leq P_+,0\leq P_-,0\leq X_4$ in $F_1$ case and $0\leq -P_+,0\leq -P_-,0\leq X_4$
$$
F_1=-{\pi^2|X_4|\over 8\mu\sin^2 ({\mu\pi\over 2})(|P_+|-|P_-|}[Y_{-\mu+1}(z_+)+Y_{\mu-1}(z_+))(Y_{-(\mu+1)}(z_-)+Y_{(\mu+1})(z_-))-
$$
$$
(Y_{-(\mu+1)}(z_-)+Y_{(\mu+1)}(z_-)(Y_{-(\mu+1)}(z_+)+Y_{(\mu+1})(z_+))]({|P_{+}P_{-}|\over |X^2_4|})^{\mu+1\over 2}
$$
$$
F_3=-{\pi^2|X_4|\over 8\mu\cos^2 ({\mu\pi\over 2})(|P_+|-|P_-|}[I_{-\mu+1}(z_+)-I_{\mu-1}(z_+))(I_{-(\mu+1)}(z_-)-I_{(\mu+1})(z_-))-
$$
$$
(I_{-(\mu+1)}(z_-)-I_{(\mu+1)}(z_-)(I_{-(\mu+1)}(z_+)-I_{(\mu+1})(z_+))]({|P_{+}P_{-}|\over |X^2_4|})^{\mu+1\over 2}
$$
All calculations above was done in coordinate system in which $X=(X_4,0,0,0),\quad P=(P_4,P_3,0,0)$. Thus $ (PX)=P_4X_4,(XX)=X_4^2,(PP)=P_+P_-,
P^2X^2-(PX)^2=-P_3^2X_4^2,X_4P_+=(PX)\pm\sqrt{(PX)^2-P^2X^2},X_4P_-=(PX)\mp\sqrt{(PX)^2-P^2X^2}$
In formulae for $F_i$ functions the Bessel function are presented only in two combination
$$
\tilde K_{\sigma}(x)\equiv {I_{-\sigma}(x)-I_{\sigma}(x)\over \sin ({\sigma\pi\over 2})},\quad \tilde L_{\sigma}(x)\equiv {Y_{-\sigma}(x)+Y_{(\sigma}(x)\over \cos ({\sigma\pi\over 2})}
$$
both invariant with respect transformation $\sigma \to -\sigma$

\subsubsection{$X^2\leq 0$}

In this case function under integral in connection with (\ref{7}) is
$$
-{1\over P_4} [\sin {X_3\over 2}(y-{P_+\over X_3y})\cos {X_3\over 2}(x+{P_-\over X_3x})-\sin {X_3\over 2}(y+{P_-\over X_3y})\cos {X_3\over 2}(x-{P_+\over X_3x})]
$$
$$
{2\over P_--P_+}[\sin {X_3\over 2}(x-{P_+\over X_3x})\cos {X_3\over 2}(y+{P_-\over X_3y})-\sin {X_3\over 2}(x+{P_-\over X_3x})\cos {X_3\over 2}(y-{P_+\over X_3y})]
 $$
The above function is invariant with respect to following exchange of variables $P_4\to -P_4,P_+\to -P_-,P_-\to -P_+$ and $ X_3\to -X_3, P_+\to P_-,P_-\to P_+$.

As in previous subsubsection integral is divided on 3 parts
$$
[\theta(X_3)\theta(P_+)\theta(-P_-)+\theta(-X_3)\theta(P_-)\theta(-P_+)]f_1=
$$
$$
\theta(-P_-P_+)[\theta(X_3)\theta(P_3)+\theta(-X_3)\theta(-P_3))]F_1=\theta(-P_-P_+)\theta(X_3P_3)F_2(X_4\to X_3)
$$
$$
[\theta(X_3)\theta(-P_+)\theta(P_-)+\theta(-X_3)\theta(-P_+)\theta(P_+)]f_2=-
$$
$$
\theta(-P_-P_+)[\theta(X_3)\theta(-P_3)+\theta(-X_3)\theta(P_3))]f_2=\theta(-P_-P_+)\theta(-X_3P_3)F_1(X_4\to X_3)
$$
$$
[\theta(X_3(\theta(P_+)\theta(P_-)+\theta(-P_-)\theta(-P_+))+\theta(-X_3)(\theta(P_+)\theta(P_-)+\theta(-P_+)\theta(-P_-))]f_3=
\theta(P_+P_-)_3(X_4\to X_3)
$$
All calculations in this subsubsection are in the system of coordinates with $X=(0,X_3,0,0),P=(P_4,P_3,0,0)$ and thus $X^2=-X_3^2,P^2=P_-P_+,(PX)=-P_3X_3,P^2 X^2-(PX)^2=-P_4^2 X_3^2,P_4 X_3=\pm\sqrt{(PX)^2-P^2 X^2},z_{\pm}^2=|X_3 P_{\pm}|=|P_4X_3\pm X_3P_3=|=|
\pm\sqrt{(PX)^2-P^2 X^2}\mp (XP)|=|\pm\sqrt{(PX)^2-P^2 X^2}\mp (XP)|$. $z_+^2=|\mp\sqrt{(PX)^2-P^2 X^2}+(XP)|,z_-^2=|\pm\sqrt{(PX)^2-P^2 X^2}+(XP)|$
This is exactly the same result as in previous subsubsection Thus finally summarizing all results we obtain for wave function of modified world
$$
<P|X>=\theta(X^2 P^2)\theta((XP))F_1+\theta(X^2 P^2)\theta(-(XP))F_2+\theta(-(X^2 P^2))F_3
$$
Some conclusion to this section. By the reason of cumbersome of all calculation author can't exclude misprints and mistakes. But if results are correct author absolutely sure that there exist some more simple way for it's obtaining

\section{Outlook}

Let us compare result of present paper with the  description arising in the usual Poinc\'are-Minkowsky world. There wave function  is equal $<X|P>=e^{i(XP)\over h}$. $|<X|P>|^2=1$. This means that in this world limitation on dynamics is very pure. The object with dynamical characteristic $P$ has equal probability to exist in arbitrary point  $X$, and to change this situation it is necessary to introduce some
outside interaction (potential). In the classical domain moving (without outside interaction) only along the straight lines with constant velocity infinitely long time. In this sense Poinca\'re-Minkowski world is not dynamical one.
In contradiction to it considered above modified world allow the classical consideration only on the limited time interval. The objects Classical observed as a classical one may arise in some space-time point and disappear  in another one. Of course interaction between the objects present but  whose influence is more important interaction with other objects or with space-time manifold is the question..
Thus, there arises a hypothesis that the star sky is the r4esult of not only gravitation interaction of its objects but also the dynamical 
nature of the modified space-time manifold. In this sense modified space-time manifold may be considered as alternative to general relativity.

Generalized quantum space-time of the present paper is only simple example of much more grande problem.

Author hope in the nearest future consider modified space-time manifold with four commutative coordinates but with de-Sitter group of motion. This correspond to the $M^2\to \infty$ in the algebra of the introduction of this paper (\ref{2}).
In this case in the physics of elementary particles the change should be no less revolutionary,
because symmetry does not know anything about the distances, and replacing a  non-semi-simple Poinc\'are algebra by a semi-simple de-Sitter one may lead to unexpected consequences.

And at last the proposed (model) theory of the Universe is not time invariant \cite{LezKhr73} and this circumstance also may have unexpected repercussions both on microscopic and on macroscopic scales of observation.

\section{Acknowledgments}

The author would like  to thanks J.Uruchyrty for many fruitful discussions, D.B.Fairlie for absolutely negative estimation of the main idea of the paper above after which author have decided to publish it. My best thanks to A.V.Razumov for the help in preparation manuscript to publication.

\end{document}